\documentclass[12pt,english]{article}
\usepackage{babel}
\usepackage{epsfig}
\usepackage{amsmath}
\usepackage{amssymb} 
\usepackage[small]{caption2} 
\usepackage{fleqn} 
\usepackage{graphicx} 
\usepackage[small,loose]{subfigure}  
\usepackage{cite} 
\advance \headheight by 3.0truept       
\setcaptionwidth{.85\textwidth}

\newcommand{\CenterObject}[1]{\ensuremath{\vcenter{\hbox{#1}}}}

\makeatother

\begin{document}
\title{
\begin{flushright}
{\normalsize DESY 05-086}
\end{flushright}
\vspace{1cm}
{\bf A Stringy Solution to the FCNC Problem\footnote{Based on a talk given at
Planck'05, 23-28 May 2005, ICTP, Trieste, Italy}}\\[0.8cm]}
\author{%
\textbf{Oleg~Lebedev }\\[0.4cm]
{\normalsize\textit{$$Deutsches Elektronen-Synchrotron DESY, 
22603 Hamburg, Germany}}}
\maketitle 
\hrule
\abstract{
The solution to the supersymmetric FCNC problem may not have a simple
field theory interpretation. 
Here I advocate a possibility based on
string selection rules. The requirement of non--trivial Yukawa structures
 restricts the form of the soft terms such that the FCNC problem
often does not arise.
}
\vspace{0.5cm}
\hrule
\vspace{1.5cm}

{\bf 1. Introduction}\\ \

Generically, supersymmetric contributions to flavor changing neutral currents
(FCNC) are orders of magnitude too large, which is known as the 
supersymmetric FCNC problem \cite{Ellis:1981ts}. 
This problem appears in models with a light
($<$ 1 TeV) SUSY spectrum and arbitrary flavor structure of the soft terms.
Simple field theory arguments suggest the following  common remedies for the problem:

$\bullet$ universality

$\bullet$ decoupling 

$\bullet$ alignment \\
The first possibility implies that the soft terms are flavor--blind, the second
-- that the first two sfermion generations are very heavy, and the third --
that the soft and SM flavor structures are aligned. Any of these options
leads to sufficient suppression of the unwanted effects, yet it is quite
nontrivial to embed these mechanisms in string theory. For example, 
universality of the soft terms would imply a special point in  moduli space
and one would have to argue that the Universe rests precisely there.
The decoupling option assumes a rather peculiar SUSY spectrum and so on. 
This is not to say that these solutions cannot be realized, rather they leave
one wondering if there could be options suggested by  string theory itself
and which may not have a simple bottom--up interpretation.
I will try to argue that such an option exists and it is based on\\ \

$\bullet$ string selection rules \\ \ \\
The basic idea is that all flavor effects are interrelated in string theory
and the FCNC problem should be analyzed in settings which allow
for realistic Yukawa matrices,
\begin{center}
       \fbox {{\rm  Yukawa couplings} $~~\leftrightarrow~~$    {\rm FCNC}}               
\end{center}
In this  framework, the form of the soft
terms is restricted and the FCNC problem  often does not arise.\\ \

{\bf 2. String Yukawa couplings }\\ \

Replication of chiral fermion families and generation of non--trivial 
flavor structures have been successfully realized in heterotic and intersecting brane models.
The fermion mass hierarchy appears naturally if
the matter fields are twisted, i.e. localized at special points in the 
compactified 6D space. The Yukawa coupling among three fields
located at three different points is exponentially suppressed by the area of the triangle
formed by these points \cite{Hamidi:1986vh}. 
Then order one variations in the area translate into a
large hierarchy among the Yukawa couplings.

In heterotic orbifold compactifications, twisted states are localized at orbifold
fixed points. Not all fixed points can couple together, they have to satisfy
string selection rules \cite{Hamidi:1986vh}. Consider
three fields $S_a$ ($a$=1,2,3) belonging to twisted sectors $\theta_a$
and localized at fixed points $f_a$. That is,
the boundary conditions for the corresponding closed string coordinates are given by
$$X_a(\tau,\sigma=\pi)=\theta_a X_a(\tau, \sigma=0) +l_a$$ 
with $l_a$ being a torus lattice vector in the conjugacy class associated with
the fixed point $f_a$.
The Yukawa couplings  $S_1 S_2 S_3$   are allowed only if (see e.g.\cite{Bailin:1999nk}) 
$$ \theta_1 \theta_2 \theta_3 = {\bf 1} $$
and 
$$ ({\bf 1}-\theta_1)f_1 + ({\bf 1}-\theta_2)f_2 + ({\bf 1}-\theta_3)f_3=0~. $$
This is required by  the existence of classical solutions to
the string equations of motion  with proper boundary conditions (instantons),
which are responsible for  correlators of the twist fields.
These constraints can be derived from the monodromy condition
 $\int _C dz ~\partial X + \int_C d \bar z ~ \bar \partial X =0$
with $z=\exp({\tau + i \sigma})$, $X(z,\bar z)$ being a classical solution to the
string equations of motion in the presence of the twist fields, and  
 $C$ is a contour encircling the positions of all three twist fields.
If this condition is not satisfied, the correlator of the twist fields vanishes
and the corresponding Yukawa coupling is zero.  
Physically, the reason for this selection rule is that the strings should
have proper boundary conditions to be able to join or split, i.e. participate
in interactions.

For the allowed Yukawa couplings, the result depends on the T-modulus:
$$  Y_{_{123}} \sim {\rm e}^{-\alpha_{_{123}} T}   $$
with  $\alpha_{_{123}}  \leq {\cal O}(1)$. If all $S_a$ are located at the same fixed point,
$\alpha_{_{123}}=0$ and the coupling is order one, otherwise it is exponentially suppressed.
This is the origin of the fermion mass hierarchy.
As explicit calculations show, semi--realistic Yukawa matrices can be produced if
{\it different generations} of one  matter field  belong to the 
{\it same twisted sector}.  To illustrate this point, let us consider a few examples.

{\bf (1). $Z_3$.} This orbifold has nine moduli associated with the sizes of the $T^2$-tori
and angles between them. If
the matter fields are twisted, the Yukawa couplings are functions of these moduli
which provides enough freedom to fit the fermion 
masses\footnote{The quark mixings can be generated by non--renormalizable
operators \cite{Casas:1989qx}  or by introducing 3 generations of Higgses \cite{Abel:2002ih}. }
 \cite{Casas:1989qx}. 
There is only one twisted sector $\theta$ 
and the allowed coupling is of the type $\theta\theta\theta$. 
Note that if one assumes that some of the fields are untwisted, the Yukawa couplings
would essentially be zero or one. Thus, the desired moduli dependence will be lost.  

{\bf (2). $Z_4$.} There are two twisted sectors $\theta, \theta^2$ and 4 moduli which enter
the Yukawa couplings $\theta\theta\theta^2$.
A satisfactory fit to the quark masses
can be obtained if different generations belong to the same twisted sector,
e.g. the quark, lepton and Higgs  doublets are in $\theta$, while the singlets are in $\theta^2$  
\cite{Casas:1992zt}. If one were to place, for instance, one generation of the quarks doublets
in $\theta^2$, the corresponding Yukawa coupling $\theta \theta^2 \theta^2$ would be
prohibited and the success of the fit would be lost.

{\bf (3). $Z_6$.}  This orbifold has three twisted sectors $\theta,\theta^2,\theta^3$ 
and allows for flavor non--diagonal Yukawa couplings $\theta\theta^2 \theta^3$.
A satisfactory  fit to the quark masses and mixings is obtained for
the Higgses in $\theta$, the doublets in $\theta^2$ and the singlets in $\theta^3$ \cite{Ko:2004ic}.
This is also required by the presence of physical CP violation at the 
renormalizable level, i.e.
a non--zero Jarlskog invariant \cite{Lebedev:2001qg}.
If some of the generations were reassigned to a different twisted sector, the Yukawa matrices
would contain many zeros and the Jarlskog invariant as well as some mixing angles
would vanish.

There are, of course, many more similar examples. The main message is that a satisfactory
fit to the quark masses or mixings at the renormalizable level  appears to require 
that different fermion generations belong to the same twisted sector\footnote{
This is also favored by the mechanisms of family replication in the
heterotic string (see e.g. \cite{Ibanez:1987sn},\cite{Buchmuller:2004hv}).}.

\begin{figure}[t]
 \centerline{\CenterObject{\includegraphics[width=5.0cm]{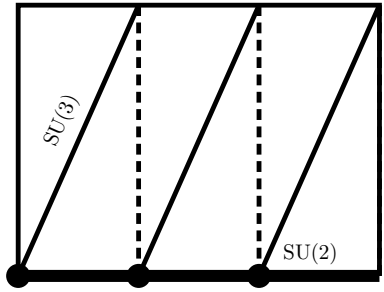}}}
\vspace*{0.3cm}
 \caption{
Family replication in intersecting brane models. 
The picture represents a $T^2$ torus  with multiple brane intersections
corresponding to three generations of the quark doublets.  }
\label{f1} 
\end{figure}

This result also applies to semirealstic intersecting brane models \cite{Blumenhagen:2000wh}. 
The chiral
fermions are localized at intersections of the branes which support the 
Standard Model gauge groups. Different generations correspond to different
intersections of the same branes (Fig.\ref{f1}). Their Yukawa couplings are again exponentially
suppressed by the area of the triangle formed by the Higgs, the left--handed  and the 
right--handed  quark vertices\footnote{For a discussion of the selection rules,
see \cite{Higaki:2005ie}.}. Realistic quark masses and mixings can be
obtained in such settings \cite{Cremades:2003qj}. What is important is that the existing mechanism of 
family replication implies that different generations differ only in their
distances to the location of the Higgs, but are the same in other aspects.
In particular, the angles between the branes are the same at different intersections (Fig.\ref{f1})
leading to the same K\"ahler potential. Thus we reach the same conclusion
as in the heterotic string case. \\ \

{\bf 3. Soft terms and FCNC}\\ \

What are the implications of the above conclusion  for the soft terms ?
To answer this question, one has to analyze  the K\"ahler potential,
\begin{equation}
K= \hat K + K_{\bar \alpha \beta} \phi^{* \bar \alpha} \phi^\beta +...,
\end{equation}
where $\phi^\alpha$ are matter fields.
In the heterotic string, twisted states have the following K\"ahler metric,
\begin{equation}
K_{\bar \alpha \beta}= \delta_{\bar \alpha \beta} (T+\bar T)^{n_\alpha}\;,
\label{k1}
\end{equation}
where $T$ is an overall modulus and $n_\alpha$ is a modular weight. 
Diagonality of the K\"ahler metric is enforced by the 
space group selection rules since different fixed points have
different quantum numbers with respect to this group.
The modular weights are determined by the twisted sector the state belongs to.
For non--oscillator states, 
the modular weight only depends on whether
the corresponding twist rotates all three complex planes 
in the compactified 6D space 
or just two of them \cite{Ibanez:1992hc}:
\begin{eqnarray}
&& n=-1 \;, ~~{\rm untwisted} \nonumber\\
&& n=-2 \;, ~~{\rm twisted ~with~3~planes~rotated} \nonumber\\
&& n=-1 \;, ~~{\rm twisted ~with~2~planes~rotated} \;. \label{weight}
\end{eqnarray}
In semirealistic models,
quark fields usually correspond to non--oscillator states,
so oscillators will not be discussed further.

In intersecting brane models, a similar expression holds and the role of
the modular weights is played by the angles between the relevant branes $\pi\nu_\alpha$
\cite{Lust:2004cx}:
\begin{equation}
K_{\bar \alpha \beta} \propto \delta_{\bar \alpha \beta} (T+\bar T)^{-\nu_\alpha}\;.
\label{k2}
\end{equation}

What these expressions tell us is that (1) the K\"ahler potential is 
{\it flavor--diagonal}, (2) the diagonal entries are $equal$ if different
generations belong to the same twisted sector. This is intuitively clear from
the intersecting brane picture (Fig.\ref{f1}): localized fields couple only
to themselves in the K\"ahler potential (to leading order) and
these couplings are generation--independent since  
the angles between the branes are the same.

The soft terms relevant to the FCNC problem depend on the 
K\"ahler potential and the Yukawa couplings $Y_{\alpha\beta\gamma}
\equiv Y_{Q_{_{L_i}} Q_{_{R_j}} H}~$ \cite{Soni:1983rm},
\begin{eqnarray}
&&m_\alpha^2=(m_{3/2}^2 +V_0) -\bar F^{\bar m} F^n \partial_{\bar m}
\partial_n \log K_\alpha \;,\label{soft}\\
&&A_{\alpha \beta \gamma}=F^m \left[     
\hat K_m + \partial_m \log Y_{\alpha \beta \gamma}- \partial_m\log
(K_\alpha K_\beta K_\gamma)   \nonumber
\right]\;,
\end{eqnarray}
where a diagonal K\"ahler metric has been assumed,
$K_{\bar \alpha \beta}=\delta_{\bar \alpha \beta} K_\beta$.
$m_\alpha^2$ are the squark masses and  $A_{\alpha \beta \gamma}$
are the trilinear soft couplings. $F^n$ is an F--term associated
with a hidden superfield $h^n$ and subscripts denote differentiation
with respect to these fields. The gravitino mass and the ``vacuum energy''
are denoted by $m_{3/2}$ and $V_0$, respectively.

\begin{figure}
\epsfig{figure=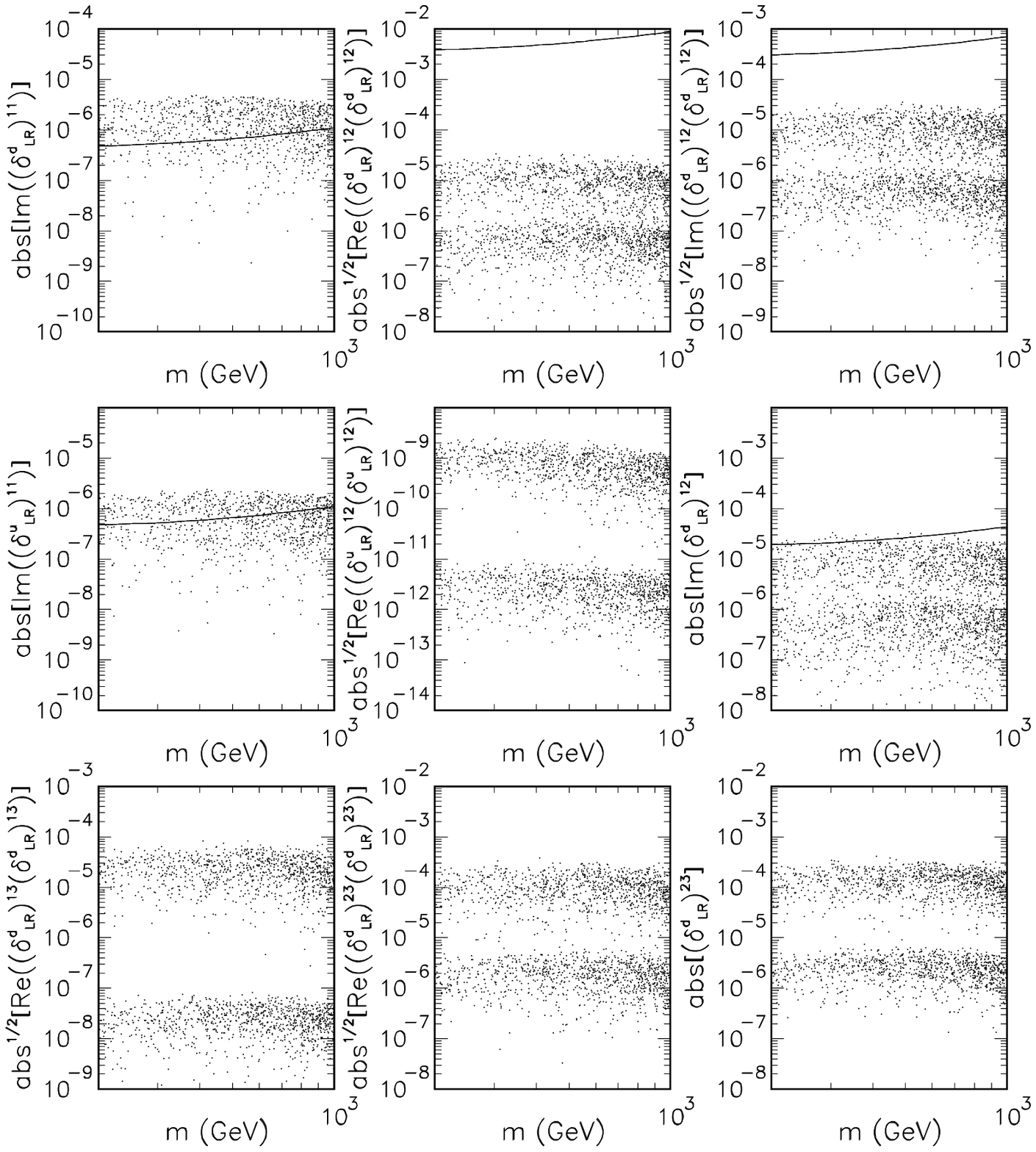,width=\linewidth} 
\vspace{1.0truecm}
\caption{\small{
Flavor violating  mass insertions  due to  non--universal
A--terms \cite{Chankowski:2005jh}. 
The SUSY parameters are chosen as
$m=2A$, $M_{1/2}=200$~GeV, $\tan\beta=15$
and order one $\tilde A_{ij}^{u,d}$ are generated randomly.
All experimental bounds (marked by a line)  on FCNC
observables  are satisfied. 
(The nEDM, which is a {\it flavor--conserving} observable, is
problematic as shown in the two top left blocks. This represents
the SUSY CP problem which is not solved by the field theory mechanisms
mentioned in the introduction.)
}}
\label{fig:lrydg2}
\end{figure}

Eqs.(\ref{k1}),(\ref{k2})  and (\ref{soft}) tell us that the squark masses
are {\it generation--independent}  
as long as different generations belong to the same twisted
sector. This means that the FCNC problem is absent. Note that
the Yukawa and the A--term flavor structures can still be complicated \cite{Abel:2001cv}. 
The FCNC induced by the A--terms are small  \cite{Chankowski:2005jh}   since the 
corresponding SUSY contributions
are suppressed by  the quark masses. This is illustrated in Fig.\ref{fig:lrydg2}.

Thus, the correlation between the Yukawa structure and the K\"ahler
potential in string models suppresses FCNC to a desired level.
The above considerations apply mostly to the quark sector. In the lepton sector,
FCNC constraints require the charged slepton masses to be universal and the A--terms
to be diagonal \cite{Chankowski:2005jh}. The latter condition 
is specific to the lepton sector
 and   implies  that the corresponding Yukawa matrix is diagonal.
This can be implemented in many orbifold models with diagonal
selection rules for renormalizable couplings, however the analysis
is obscured by the unknown origin of the neutrino masses and mixings
(see e.g. \cite{Giedt:2005vx}). \\ \

{\bf 4. Relaxing the assumptions}\\ \

The conclusion that the FCNC are suppressed is based on the analysis of  renormalizable
couplings in the superpotential.
One may wonder if non--renormalizable couplings can affect it.
There are two types of such contributions:

(i) non--renormalizable couplings are only perturbations over the renormalizable 
Yukawas

(ii) non--renormalizable couplings lead to qualitative changes such as the 
twisted sector assignment\\
In the former case the above discussion applies, whereas in the latter
the conclusions  may change.

In general, some of the Yukawa entries may only be allowed at the non--renormalizable
level. For instance, off--diagonal entries of the Yukawa matrices
in $Z_3$ are forbidden at the renormalizable level (see, however, \cite{Abel:2002ih})
and are induced by higher dimensional
operators involving singlet fields, e.g. of the type $\theta\theta\theta (\theta)^9$ 
\cite{Casas:1989qx}.
In many cases, the presence of such operators does not change the conclusion
that different generations should belong to the same twisted sector in order to
get a good fit to the quark masses and mixings.
For instance, in $Z_3$ and $Z_7$ there is only one twisted sector, so this condition 
will be satisfied as long as the fields are twisted.
Yet, it is also possible to evade this conclusion in some cases.

\begin{figure}
\epsfig{figure=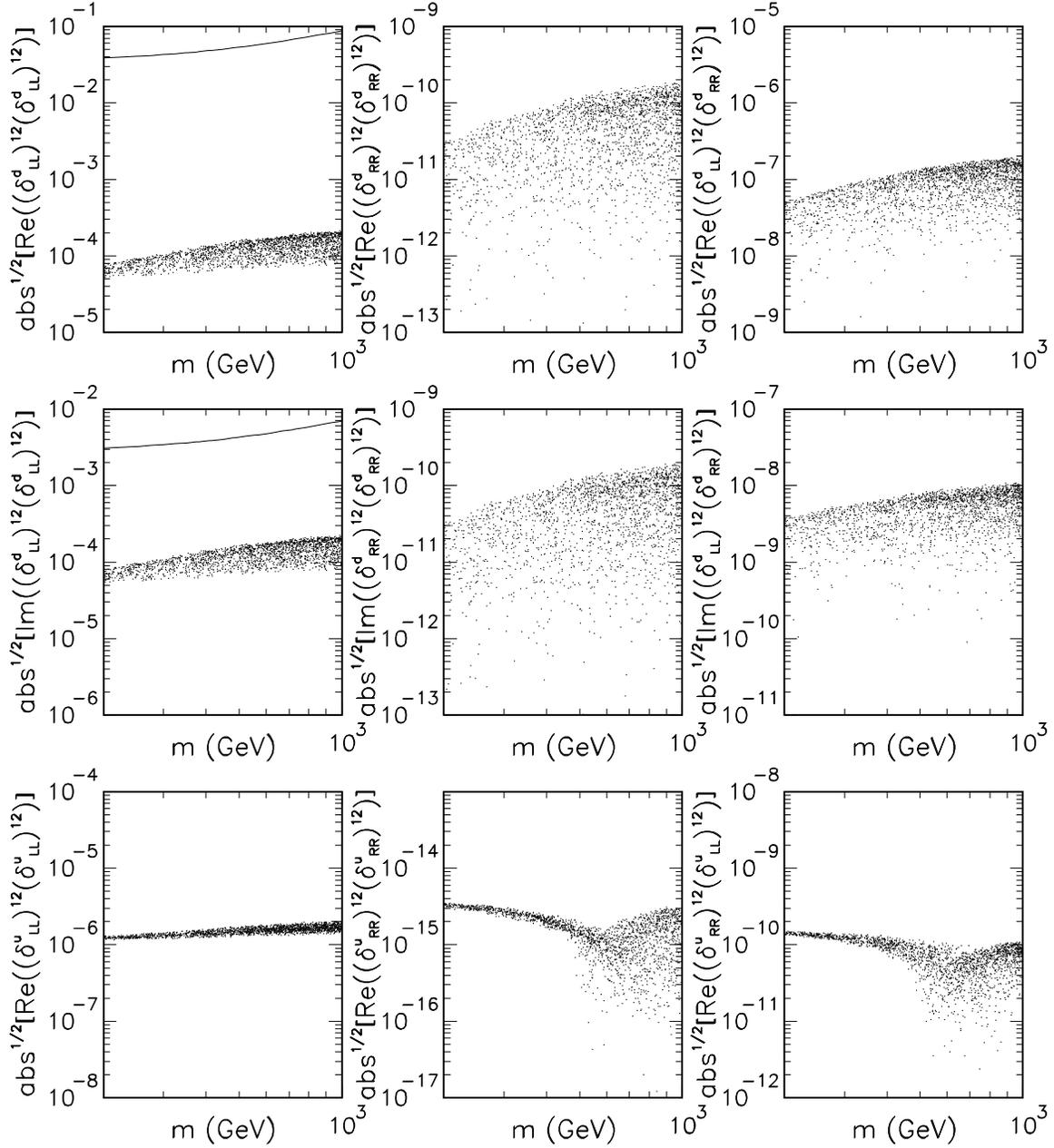,width=\linewidth} 
\vspace{1.0truecm}
\caption{\small{
Flavor violating mass insertions  due to  non--universal
soft scalar masses, $m_1=m_2\not=m_3$  \cite{Chankowski:2005jh}.  
The SUSY parameters are chosen as
$A=0$, $M_{1/2}=200$~GeV, $\tan\beta=15$
and $m_3$ is varied randomly in the range $m_1/2\div m_1$.
The Yukawa matrices are assumed to be diagonalized by 
matrices similar to the CKM one.
All experimental bounds (marked by a line)  on FCNC
observables  are satisfied. 
(The $B$-physics observables are not shown.) }}
\label{fig:scalar}
\end{figure}

Consider splitting the third family from the other two. For example, in $Z_6$ one
can place the first two families in the $\theta$ twisted sector and the third family
in the untwisted sector \cite{Kobayashi:2004ud}. In this case, the K\"ahler metric
remains diagonal, yet the modular weights  for the light and heavy families become different
(Eq.(\ref{weight})). That means, the squark masses at the GUT scale have the structure
$$ m^2_i={\rm diag}(m^2,m^2,M^2)~, $$ 
with $m^2-M^2 \sim m^2$. What is relevant for the FCNC is off--diagonal elements
of the squark mass matrix at low energies in the physical (sCKM)  basis, i.e. the basis
in which the quark masses are diagonal. To translate the above GUT boundary condition
into these quantities, one needs to (1) RG--run  $m^2_i$ to low energies, (2)
rotate  $m^2_i$ to the sCKM basis. Both of these effects are very important.
First, the gluino renormalization effect increases $m^2_i$ by about an order
of magnitude, $m^2_i \rightarrow m^2_i + 8 m_{\tilde g}^2$
\cite{Brignole:1993dj}. Second, the rotation
to the sCKM basis induces the off diagonal elements $(m^2-M^2) \epsilon$, where 
$\epsilon $ is the rotation angle. The quantity governing the size of SUSY contributions
to the FCNC is
$$ \delta \sim   {m^2-M^2 \over m^2+8 m_{\tilde g}^2} ~\epsilon \leq 10^{-3}~,\label{M}$$
with $m^2 \sim m_{\tilde g}^2$ and $\epsilon \leq 10^{-2}$. 
For hierarchical Yukawa textures and the gluino mass larger or similar
to the squark masses, 
it is usually small enough to satisfy all experimental
FCNC constraints \cite{Chankowski:2005jh}. Thus, we see that this scenario
does not suffer any significant FCNC problem (Fig.\ref{fig:scalar}).

One could also consider the possibility that all three generations are 
split,
$$ m^2_i={\rm diag}(m^2,M^2,{\cal M}^2)~. $$ 
(It would be difficult however to find convincing examples.)
In this case, the FCNC constraints are more severe, yet the  RG--dilution mechanism 
 is at work. One finds that the FCNC constraints are satisfied as long
as the rotation matrices to the physical basis are similar to the CKM
matrix  \cite{Chankowski:2005jh}. 

Thus, some of our  assumptions can be relaxed without encountering a severe FCNC
problem. What is important in this analysis  is that the K\"ahler metric stays
diagonal\footnote{Higher dimensional operators
may lead to off--diagonal terms in the K\"ahler metric.
These however are significantly suppressed.}, 
which is true in essentially all reasonable 
models (see e.g. \cite{Brignole:1995fb}).
Finally, note that string theory restricts not only the form
of the K\"ahler potential but also the ``numbers'' entering into it:
the modular weights in semirealistic models are 
 either -1 or -2  (Eq.(\ref{weight})), so it is likely that the
first two generations have the same weights purely on the statistical
basis.\\ \

{\bf 5. Conclusion}\\ \

Correlations between the structure of the Yukawa couplings and that of the 
K\"ahler potential suppress SUSY--induced FCNC in string models.
This conclusion is valid for a standard SUSY spectrum with no peculiarities,
nor does it rely on family symmetries or universality. 
It is based on string selection rules which stem from 
properties of the compactified space
in  heterotic  and intersecting
brane models.

The question that remains is what part of the analysis will survive
in a truly realistic model. Perhaps the details will change entirely,
but, as the above examples show, string theory may be clever enough
to avoid  automatically  the problems that bug phenomenologists.

{\bf Acknowledgements.} I would like to thank  P. Chankowski
and S. Pokorski for collaboration on \cite{Chankowski:2005jh}. 
I am also grateful to M. Ratz for his help with the manuscript
and to S. Abel, W. Buchm\"uller, S. Raby and O. Vives
for helpful comments.

\newpage
\newpage

\newpage
\newpage

\end{document}